# Neuronal Avalanches Imply Maximum Dynamic Range in Cortical Networks at Criticality

Abbr. Title: ***Dynamic Range Maximized During Neuronal Avalanches***

Woodrow L. Shew[1], Hongdian Yang[1,2], Thomas Petermann[1], Rajarshi Roy[2], and Dietmar Plenz[1]

[1]*Section on Critical Brain Dynamics, Laboratory of Systems Neuroscience, National Institute of Mental Health, 9000 Rockville Pike, Bethesda, Maryland 20892, USA*

[2] *Institute for Physical Science and Technology, University of Maryland, College Park, Maryland 20742, USA*

| | |
|---|---|
| Correspondence: | Dietmar Plenz, Ph.D., Section on Critical Brain Dynamics /LSN, Porter Neuroscience Research Center, National Institute of Mental Health, Rm 3A-100, 35 Convent Drive, Bethesda, MD 20892, office: (301) 402-2249, fax: (301) 480-7480; plenzd@mail.nih.gov |
| Figures/Tables: | 4/0 |
| Suppl. Material: | 8 pages / 6 Figures |
| Pages: | 20 |
| Total #words: | 4675 |
| Keywords: | neuronal avalanches, balance of excitation and inhibition, critical phenomena, input-output dynamic range, organotypic culture, multielectrode array, model |

Acknowledgement:   We thank Craig Stewart for help with preparing the cultures. This work was supported by the Intramural Research Program of the NIMH.  RR acknowledges support from DOD MURI grant (ONR N000140710734) to UMD. TP is also grateful to the Swiss NSF (Grant No. PBEL2-110211) for financial support.

**Spontaneous neuronal activity is a ubiquitous feature of cortex. Its spatiotemporal organization reflects past input and modulates future network output. Here we study whether a particular type of spontaneous activity is generated by a network that is optimized for input processing. Neuronal avalanches are a type of spontaneous activity observed in superficial cortical layers *in vitro* and *in vivo* with statistical properties expected from a network in a 'critical state'. Theory predicts that the critical state and, therefore, neuronal avalanches are optimal for input processing, but until now, this is untested in experiments. Here, we use cortex slice cultures grown on planar microelectrode arrays to demonstrate that cortical networks which generate neuronal avalanches benefit from maximized *dynamic range*, i.e. the ability to respond to the greatest range of stimuli. By changing the ratio of excitation and inhibition in the cultures, we derive a network tuning curve for stimulus processing as a function of distance from the critical state in agreement with predictions from our simulations. Our findings suggest that in the cortex, (1) balanced excitation and inhibition establishes the critical state, which maximizes the range of inputs that can be processed and (2) spontaneous activity and input processing are unified in the context of critical phenomena.**

**Introduction**

The cortex displays ongoing activity that persists even in the absence of any obvious stimulus or motor output.  Increasingly, evidence shows that ongoing activity is intricately linked to stimulus-evoked activity.   For example, orientation maps constructed from ongoing neuronal activity in the anesthetized cat match those based on visual responses (Tsodyks et al., 1999;Kenet et al., 2003).  Spatiotemporal correlations of spikes in the visual cortex are similar when the awake animal is simply sitting in darkness or observing natural scenes (Fiser et al., 2004).  Likewise, population responses to auditory and somatosensory stimuli fall within the repertoire of observed spontaneous events (Luczak et al., 2009).  Moment to moment, ongoing activity contributes to the large variability observed in stimulus responses (Arieli et al., 1996;Kisley and Gerstein, 1999;Azouz and Gray, 1999), while being only weakly modulated by stimulus presentation (Fiser et al., 2004).  On longer timescales, the organization of spontaneous activity is thought to reflect past inputs and influence future network responses (Ohl et al., 2001;Yao et al., 2007).  Such interplay between spontaneous and stimulus-evoked activity raises the question whether there is a particular type of ongoing activity that maintains optimized stimulus processing in the network.

Here we focus on neuronal avalanches, a type of spontaneous activity that has been observed in superficial layers of cortex *in vivo* and *in vitro* (Beggs and Plenz, 2003;Plenz and Thiagarajan, 2007;Gireesh and Plenz, 2008;Petermann et al., 2009). Neuronal avalanches consist of bursts of elevated population activity, correlated in space and time, that are distinguished by a particular statistical character:  activity clusters of size *s* occur with probability $P(s) \sim s^{\alpha}$, i.e. a power law with exponent α = -1.5. Neuronal avalanches have several additional key properties: (*i*) they arise during development when superficial layers form *in vitro* and *in vivo* (Gireesh and Plenz, 2008), (*ii*) they are homeostatically maintained for weeks in isolated cortex without any input (Stewart and Plenz, 2007), (*iii*) they constitute the dominant form of ongoing cortical activity in the awake monkey (Petermann et al., 2009), and (*iv*) their pharmacological regulation is characterized by an inverted-U profile of NMDA/Dopamine-D1 receptor interaction and intact fast inhibitory transmission (Stewart and Plenz, 2006;Gireesh and Plenz, 2008;Beggs and Plenz, 2003).

Neuronal avalanches are similar to the dynamics of other systems poised at the boundary of order and disorder; more precisely, we refer here to systems operating in a *critical state* (Bak and Paczuski, 1995;Jensen, 1998;Stanley, 1971).  Importantly, simulations and theory predict that neuronal networks in the critical state optimize several aspects of information processing including (1) the range of stimulus intensities that can be processed, i.e. *dynamic range* (Kinouchi and Copelli, 2006) and (2) the amount of information that can be transferred (Beggs and Plenz, 2003;Tanaka et al., 2009).  Until now, experimental support of these predictions has been lacking.  Here, we demonstrate that *in vitro* cortical networks have maximum dynamic range when spontaneous activity takes the form of neuronal avalanches.  By systematically changing excitation and inhibition, we obtain a tuning curve for stimulus processing in cortical networks, with peak performance found under conditions which generate neuronal avalanche activity.

**Material and Methods**
*Preparation of organotypic cultures on the microelectrode arrays (MEA).* Coronal slices from rat somatosensory cortex (350 μm thick) and the midbrain (VTA; 500 μm thick) were taken from newborn rats (postnatal day 0–2; Sprague–Dawley) and cultured on a poly-D-lysine-coated 8x8 MEA (MultiChannelSystems; 30 μm electrode diameter; 200 μm interelectrode distance).  In the organotypic slice, both deep cortical layer and superficial layers develop (Götz and Bolz, 1992;Plenz and Kitai, 1996) and the development of neuronal avalanche activity in this *in vitro* co-culture during the second week, when superficial layers mature, parallels that observed *in vivo* (Gireesh and Plenz, 2008).  In short, a sterile, closeable chamber was attached to the MEA, which allowed for cultivation and repeated recording from single co-cultures for many weeks. After plasma/thrombin-based adhesion of the tissue to the MEA, 600 μl of standard culture medium was added (50% basal medium, 25% Hanks' balanced salt solution, 25% horse serum; all Sigma-Aldrich) and the MEAs were affixed to a slowly rocking tray (±65° rocking angle, 200s rocking period) inside a custom-built incubator at 35.5 ± 0.5°C (Stewart and Plenz, 2007).

*Recording of spontaneous activity.* Individual MEAs were attached to a recording head stage inside an incubator (MEA1060 w/ blanking circuit; x1200 gain; bandwidth 1 – 3000 Hz; 12 bit A/D in range 0 – 4096 mV; MultiChannel Systems, Inc.).  The stability of the recording setup allowed for repeated measurements under sterile conditions from single cultures for weeks.  The local field potential (LFP; 4 kHz sampling rate; measured against common reference electrode inside the bath) was obtained from 1 hr of continuous recordings of extracellular activity, and later low-pass filtered with a cutoff at 100 Hz (phase neutral, $4^{th}$ order Butterworth).  To establish a correlation between the LFP and neuronal spiking activity, in n = 5 cultures, extracellular activity was recorded for 15 min at 25 kHz.  In addition to extracting the LFP, the extracellular signal was filtered in the frequency band 300 – 3000 Hz and up to 78 single units were identified per culture using a combination of threshold detection and PCA-based spike sorting (Offline sorter; Plexon Inc.).

*Stimulus-evoked activity.* Immediately following each 1 hr recording of spontaneous activity, stimulus-evoked activity was measured.  Stimuli were applied at 5 second intervals at one single electrode, which was located approximately at the center of the culture within superficial cortical layers.  The stimulus consisted of a current-controlled, single shock with a bipolar square waveform: 50 μs with amplitude $S$ followed by 100 μs with amplitude $+S/2$, where $S$ was varied between 6 and 200 μA. We tested two sets of stimulus amplitudes, one with finer resolution ($S$ = 10 to 200 μA in steps of 10 μA) and the other with coarser resolution ($S$ = 6, 12, 24, 50, 65, 80, 100, 150, 200 μA).  Results were not significantly different for the two stimulus protocols.  Each stimulus level was repeated 40 times in pseudo-randomized order resulting in a recording total duration of 2,000 (coarse) or 4,000 s (fine).  The neural response to each stimulus was recorded using all electrodes except for the stimulation electrode during the 500 ms following the stimulus.  A blanking circuit, which disconnected the recording amplifiers during the stimulus, significantly reduced stimulus artifacts (Multichannel systems).  Sample rate and filtering was identical to that used for spontaneous activity recordings.

*Pharmacology.* Bath application of antagonists for fast glutamatergic or GABAergic synaptic transmission was used to change ratios of excitation to inhibition (E/I). The normal (no-drug) followed by a drug condition for each culture were typically studied within a short time (~3 hrs) to minimize potential non-stationarities during development. Stock solutions were prepared for the $GABA_A$ receptor antagonist picrotoxin (PTX), the N-methyl-D-aspartic acid (NMDA) receptor antagonist (2R)-amino-5-phosphonovaleric acid (AP5), and the α-amino-3-hydroxyl-5-methyl-4-isoxazole-propionate (AMPA) receptor antagonist 6,7-Dinitroquinoxaline-2,3-dione (DNQX). Working solutions were obtained by adding 6 μl of these stock solutions to 600 μl of culture medium in the MEA chamber to reach the following final drug concentrations: (in μM) 5 PTX, 20 AP5, 10 AP5 + 0.5 DNQX, 20 AP5 + 1 DNQX. After measurement under each drug condition, the culture was washed by replacing the culture medium with 300 μl of conditioned medium mixed with 300 μl of fresh, unconditioned medium. Most cultures recovered to the critical state within about 24 hrs. Conditioned medium was collected from the same culture the day before drug application.

*Quantifying burst size and response to stimulus.* For each electrode, we identified negative peaks in the LFP (nLFP) that were more negative than –4 SD of the electrode noise. We then identified a spatiotemporal cluster of nLFPs on the array as a group of consecutive nLFPs each separated by less than a time $\tau$ (Beggs and Plenz, 2003). The threshold $\tau$ was chosen to be greater than the short timescale of inter-peak intervals within a cluster, but less than the longer timescale of inter-cluster quiescent periods ($\tau$ = 86±71 ms for all cultures; see also Suppl. Mat.). Results were robust for a large range in the choice of $\tau$ (data not shown). The size $s$ of a cluster was quantified as the absolute sum of all nLFP amplitudes within a cluster. Similarly, the size $R$ of an evoked response was quantified as the absolute sum of nLFPs within 500 ms following a stimulus.

*Definition of* κ. We previously demonstrated that for neuronal avalanches, the probability density function (PDF) of cluster size $s$ follows a power-law with slope $\alpha = -3/2$ (Beggs and Plenz, 2003)(Fig. 2A). Thus, the corresponding cumulative density function (CDF) for avalanche sizes, $F^{NA}(\beta)$, which specifies the fraction of measured cluster sizes $s < \beta$, is a –1/2 power-law function, $F^{NA}(\beta) = \left(1 - \sqrt{l/L}\right)^{-1}\left(1 - \sqrt{l/\beta}\right)$ for $l<s<L$. Here we define a novel nonparametric measure, κ, to quantify the difference between an experimental cluster size CDF, $F(\beta)$, and the theoretical reference CDF, $F^{NA}(\beta)$,

$$\kappa = 1 + \frac{1}{m}\sum_{k=1}^{m}\left(F^{NA}(\beta_k) - F(\beta_k)\right), \qquad (1)$$

where the $\beta_k$ are $m$ = 10 discrete burst sizes logarithmically spaced between the minimum and maximum burst size observed in the experiments. Our use of cumulative distributions rather than the PDFs to calculate κ avoids sensitivity to binning choices, which are necessary for constructing a PDF. κ is in the same family of nonparametric comparisons of cumulative distributions as the Kolmogorov-Smirnov (KS) test and the Kuiper's test. κ more accurately measures deviation from neuronal avalanches than previously used methods (for further details see Suppl. Mat.).

*Dynamic range.* For each culture, we measured the response to a range of stimulus amplitudes. Using the response curve, $R(S)$, the dynamic range was defined as

$$\Delta = 10\log_{10}(S_{max}/S_{min}), \tag{2}$$

where $S_{max}$ and $S_{min}$ were the stimulation values leading to 90% and 10% of the range of $R$ respectively.

*Model.* The model consisted of $N$ all-to-all coupled, binary-state neurons ($N$ = 250, 500, and 1000) and the following dynamical rules: If neuron $j$ spiked at time $t$ (i.e. $s_j(t) = 1$), then postsynaptic neuron $i$ will spike at time $t+1$ with probability $p_{ij}$. As such, the $p_{ij}$ are a $N$x$N$ matrix representing the synaptic coupling strengths between all pairs of neurons. The $p_{ij}$ are asymmetric $p_{ij} \ne p_{ji}$, positive, time-independent, uniformly distributed random numbers with mean and standard deviation of order $N^{-1}$. If a set of neurons $J(t)$ spikes at time $t$, then the probability that neuron $i$ fires at time $t+1$ is exactly $p_{iJ}(t) = 1 - \prod_{j \in J(t)}(1 - p_{ij})$. To implement the probabilistic nature and variability of unitary synaptic efficacy, neuron $i$ actually fires at time $t+1$ only if $p_{iJ}(t) > \zeta(t)$, where $\zeta(t)$ is a random number from a uniform distribution on [0,1],

$$\begin{aligned}s_i(t+1) &= \theta[p_{iJ}(t) - \zeta(t)] \\ &= \theta\left[1 - \prod_{j \in J(t)}(1 - p_{ij}) - \zeta(t)\right],\end{aligned} \tag{3}$$

where $\theta[x]$ is the unit step function. In parallel with our experiments, we explore a range of network excitability by tuning the mean value of $p_{ij}$ from 0.75/$N$ to 1.25/$N$ in steps of 0.05/$N$ by scaling all $p_{ij}$ by a constant. In this regime of small mean $p_{ij}$, the model reduces to probabilistic integrate-and-fire, i.e. $p_{iJ} \approx \sum_{j \in J(t)} p_{ij}$ to order $N^{-2}$ accuracy. If the mean $p_{ij}$ is exactly $N^{-1}$, then, $n$ action potentials occurring at time $t$ will, on average, excite $n$ postsynaptic neurons to fire at time $t+1$, which constitutes the critical state in our model (Beggs and Plenz, 2003;Kinouchi and Copelli, 2006). When mean $p_{ij}$ is larger than or less than $N^{-1}$, the system is supercritical or subcritical, respectively. We define the control parameter of the model $\sigma \equiv N^{-1} \sum_i \sum_i p_{ij}$. In the context of dynamics, σ reflects the average ratio of spiking descendants to spiking ancestors in consecutive time steps. In the critical state, σ = 1; the coupling strengths are balanced such that, on average, the number of active sites neither grows nor decays as time passes. However, it is important to realize that even though the average level of activity in the critical state is steady, the instantaneous activity level fluctuates greatly. In order to obtain response as a function of stimulus in the model, we simulated increasing stimulus amplitude $S$ by increasing the number of initially activated neurons ($S$ = 1, 2, 4, 16, 32, 64, 128 initially active neurons). Finally, we note that our model is very similar to $N - 1$ dimensional directed percolation (Buice and Cowan, 2007). Therefore, at high dimension (N>5) and weak coupling it is expected that the model behaves as a branching process. In this context, σ is the branching parameter and the –3/2 power-law is predicted in the critical state.
To test for statistical differences between groups, a one-way ANOVA followed by a Tukey post hoc test was used.

**Results**

Cortex-VTA co-cultures from rat (n = 16), which closely parallel *in vivo* differentiation and maturation of cortical superficial layers (Gireesh and Plenz, 2008), were grown on 8x8 integrated planar micro-electrode arrays (Fig. 1A). Local field potentials (LFP) were recorded after superficial layer differentiation (>10 DIV) and analyzed to extract spatiotemporal clusters of negative LFP deflections (nLFPs; n = 47 experiments). Extracellular unit activity recorded simultaneously with the LFP revealed that sizes of nLFP clusters correlated with the level of suprathreshold neuronal activity in the network (Fig.1B; R =0.84 ± 0.13, mean ± SD; n = 5 cultures). For each experimental condition, we first measured spontaneous activity (Fig. 1C) and quantified the deviation of the observed spontaneous network dynamics from neuronal avalanche dynamics by calculating κ (Fig. 2A; see M&M). In a second step, we measured the input/output dynamic range Δ of the cultured network based on its response to a range of stimulus amplitudes (Fig. 1D; Fig. 3A). These measurements were carried out under normal conditions and repeated after changing the ratio of excitation and inhibition through bath application of the antagonists PTX or AP5/DNQX.

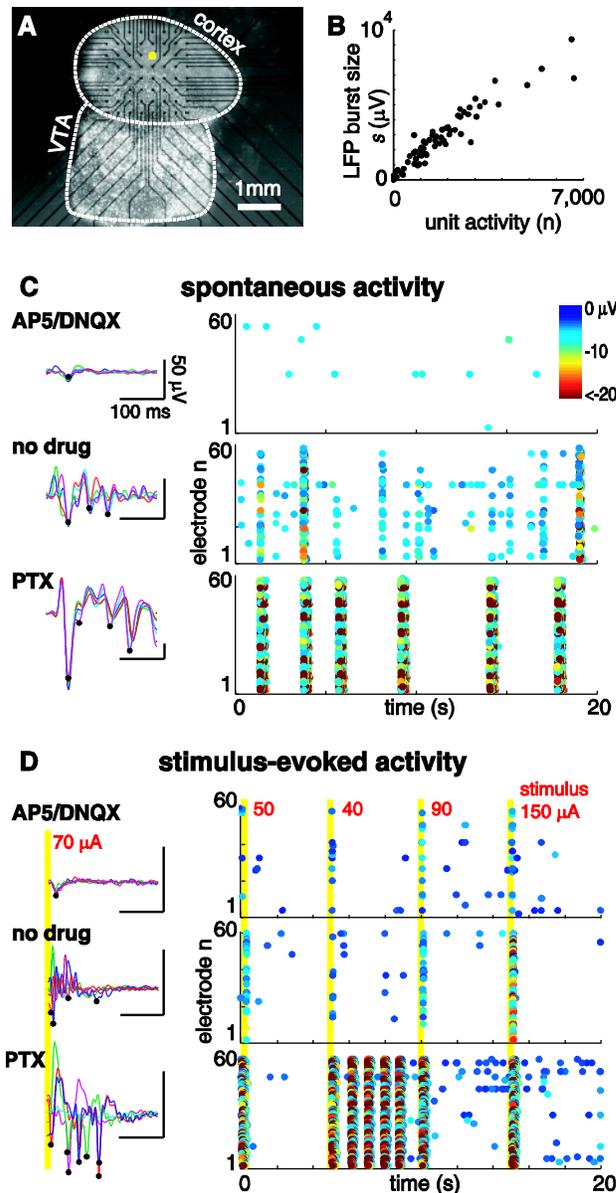

**FIGURE 1. Measuring spontaneous and stimulus-evoked activity from cortical networks.** *A,* Light-microscopic image of a somatosensory cortex and dopaminergic midbrain region (VTA) coronal slice cultured on a 60 channel microelectrode array. *Yellow dot*: stimulation site. *Black dots*: recording sites. *B,* The number of extracellular spikes on the array correlates with the size of simultaneously recorded nLFP burst (R=0.84±0.13; single culture). Each point represents total number of spikes versus the corresponding single spontaneous nLFP burst size. *C,* Example recordings of spontaneous LFP fluctuations (*left*) and nLFP rasters (*right*) for three drug conditions (*top* – AP5/DNQX, *middle* – no drug, *bottom* – PTX.) *D,* Examples of LFP evoked by 70 μA stimulus (*left*) and rasters recorded during the application of four stimuli of amplitudes 50, 40, 90, 150 μA (stimulus time marked by yellow line) (*right*) for three drug conditions. For both spontaneous (C) and stimulus-evoked (D) activity AP5/DNQX (PTX)

typically results in reduced (increased) amplitude LFP events and with lesser (greater) spatial extent. In C and D, the black dots on the LFP traces indicate nLFP events, raster point color indicates nLFP amplitude, and all scale bars (*left*) represent 50 μV, 100 ms.

*Quantifiying the cortical network state based on κ*

Figure 2A (*left*) shows experimental cluster size PDFs obtained from three cultures under normal, unperturbed conditions and in the presence of PTX or AP5/DNQX respectively. Under normal conditions, cultures revealed a PDF close to –3/2 power-law as predicted for neuronal avalanches (Fig. 2A, *black*). In the presence of PTX, however, the PDF is bi-modal revealing a high likelihood for small and large activity clusters, but a decreased probability of medium-sized clusters (Fig. 2A; *red*). In contrast, bath-application of AP5/DNQX reduces large clusters resulting in mostly small clusters (Fig. 2A; *blue*). These differences in PDFs are robustly assessed using the corresponding CDFs (Fig. 2A, *right*). Reducing excitation results in a steep early rise of the CDF, while reducing inhibition results in a delayed rise of the CDF. κ robustly quantifies these observations using the difference between a measured CDF of cluster sizes and the theoretically expected reference CDF for neuronal avalanches (Fig. 2A, *right*, *gray lines*). As summarized in Fig. 2C, $\kappa \cong 1$ under normal conditions (κ = 1.14±0.01, ±SE; n = 28), κ < 1 when excitation is reduced (κ = 0.81±0.01; n = 10) and κ>1 when inhibition is reduced (κ = 1.51 ± 0.01; n = 9; $F_{(2,44)}$ = 82.7; p<0.05 for PTX and AP5/DNQX from normal).

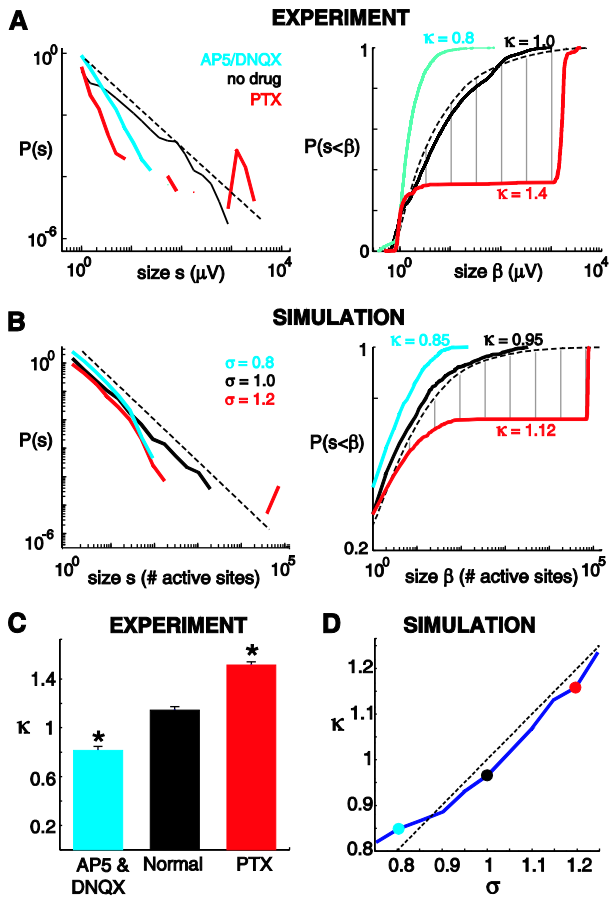

**FIGURE 2. Change in the ratio of excitation and inhibition moves cortical networks away from the critical state.** *A*, *Left*: Probability distribution functions (PDFs) of spontaneous cluster sizes for a normal, unperturbed culture (*black*), in the presence of PTX (*red*), and AP5/DNQX (*blue*). *Broken line*: -3/2 power-law. *Right*: Corresponding cumulative distribution functions (CDFs) and quantification of the network state using κ, which measures deviation from a -1/2 power-law CDF (*broken line*). *Vertical gray lines*: The 10 distances summed to compute κ shown for the example of the PTX condition (*red*). The size of a cluster *s* is defined as the sum of nLFP peak amplitudes occurring during the cluster and *P(s)* is the probability of observing a cluster of size *s*. *B*, Simulated cluster size PDFs (*left*) and corresponding CDFs (*right*) for different values of the model control parameter σ. *C*, Summary statistics of average κ values for normal (no drug), hypo-excitable (AP5/DNQX), and disinhibited (PTX) network condition (* $p < 0.05$ from normal). *D*, As shown from simulations, κ is accurately linked to σ. *Broken line*: identity. *Colored dots*: examples shown in *B*.

     This experimental strategy was paralleled using a network-level computational model of binary, integrate-and-fire neurons, in which changes in the excitation/inhibition ratio (E/I) were mimicked by tuning the parameter σ (see M&M). For σ < 1, a neuron triggers activity in less than one neuron, on average, resulting in a hypo-excitable state. Conversely, for σ > 1, one neuron excites on average more than one neuron in the near future, resulting in a hyper-excitable condition. Accordingly, for σ = 1, propagation of activity is balanced as was found experimentally for neuronal avalanches (Beggs and Plenz, 2003;Stewart and Plenz, 2007). We simulated "spontaneous" activity clusters by activating a single randomly chosen neuron and monitoring the ensuing until activity ceased or 500 time steps were executed. The total number of spikes in a cluster was taken as the cluster size. 1000 clusters were simulated at each of 11 levels of σ. In agreement with established theory, model burst size PDFs near the critical state with σ =1 fit a -3/2 power-law very closely (Fig. 2B, *right*; *black*; (Harris, 1989;Zapperi et al., 1995). Just as in the experiment, we computed κ based on CDFs of simulated spontaneous activity for different values of σ (Fig. 2B). We found that κ and σ were almost linearly related (Fig. 2D), which supports the following interpretation: In the experiments, κ ≅ 1 is close to criticality, κ < 1 identifies the subcritical regime, and κ > 1 is analog to the supercritical regime of the model.

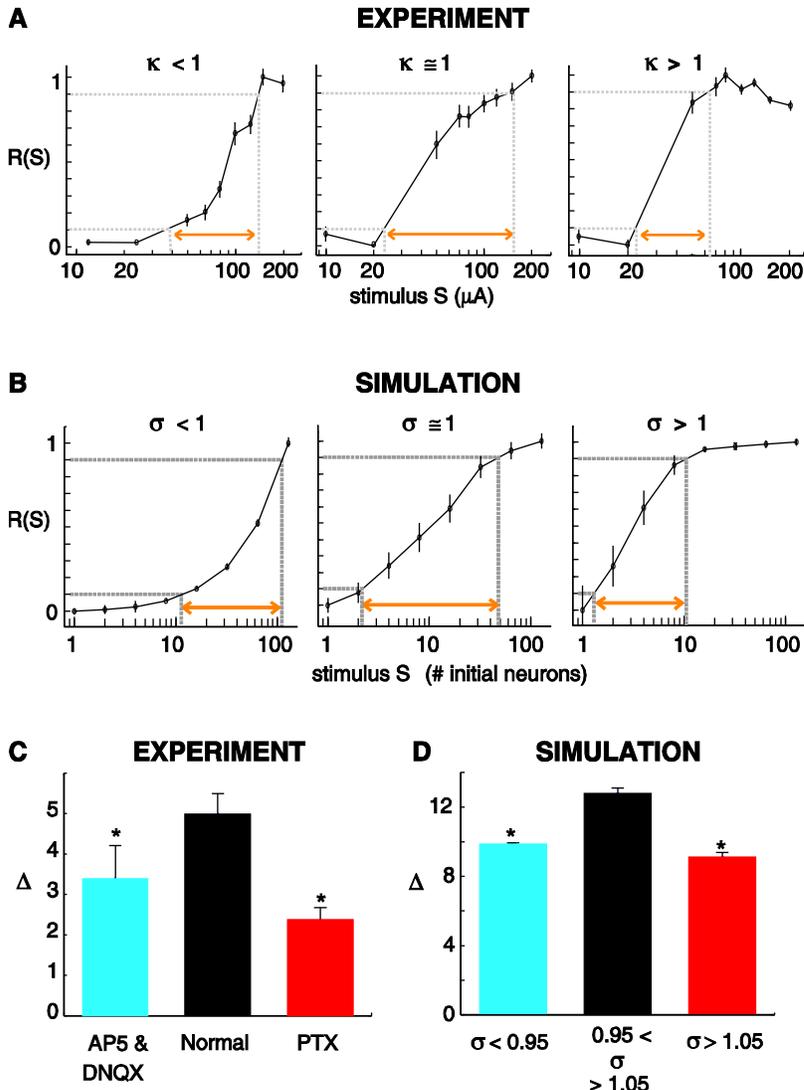

**FIGURE 3. Stimulus-response curves and dynamic range Δ.** *A,* Experimental response *R* evoked by current stimulation of amplitude *S* for three example cultures with different κ values. *Orange arrows*: Indicates range of $S_{min}$ to $S_{max}$; arrow length is proportional to Δ. Note that Δ is largest for $κ ≅ 1$. *B,* Model response curves obtained for different numbers of initially activated sites Δ is largest for $σ ≅ 1$. Like the experiment, each point is calculated from 40 stimuli at each level. *Error bars*: 1 S.E. *C,* Summary statistics for experimental values of κ under different pharmacological conditions (* $p < 0.05$ from normal). *D,* Summary statistics for model simulations and different ranges of σ (* $p < 0.05$ from $σ ≅ 1$).

*Stimulus-evoked activity and dynamic range*
After obtaining κ for a given experimental condition, we recorded the response *R* as a function of stimulus amplitude *S* (for peristimulus time histograms of *R* and different *S* see Suppl. Mat. Fig. S3). Typical response curves from experiments and simulations are shown in figure 3A and 3B respectively. We found that the shape of the response curves in the model closely matched the experimental findings. When excitatory synaptic transmission was reduced ($\kappa < 1$), the system was relatively insensitive (required a larger stimulus to evoke a given response). When inhibitory synaptic transmission was reduced ($\kappa > 1$) the system was hyper-excitable, with responses that saturate for relatively small stimuli. In the balanced E/I condition with $\kappa \cong 1$, the range of stimuli resulting in non-zero and non-saturated response was largest.

*Maximal Dynamic range at criticality, $\kappa \cong 1$*
For each response curve, we quantified the range of stimuli the network can process, i.e. the dynamic range $\Delta$ (see M&M). We found experimentally that $\Delta = 5.0 \pm 0.1$ (mean±S.E) under normal conditions, $\Delta = 2.4 \pm 0.1$ in the presence of PTX, and $\Delta = 3.4 \pm 0.3$ for AP5/DNQX (Fig. 3C, $F_{(2,44)} = 11.3$; $p<0.05$ PTX and AP5/DNQX from normal). Importantly, the dynamic range was largest in unperturbed networks, in which neuronal avalanches are most likely to occur. These results were robust for different maximal stimulus amplitudes, that is, even when the input-output response curves did not saturate for all conditions (see Supplemental Material). Similar overall changes in $\Delta$ were also found in our simulations (Fig. 3D; $F(2,195) = 820$; $p<0.05$)

We then derived a tuning curve of $\Delta$ versus $\kappa$ by combining all experimental conditions into one scatter plot (Fig. 4A). These data demonstrate that $\Delta$ is maximized and its variability is largest near $\kappa \cong 1$. These findings were closely matched by our model including observed changes in $\Delta$ as the system is pushed away from $\kappa \cong 1$, ~10dB drop (10 fold reduction in $S_{max}/S_{min}$) for a 30% change in $\kappa$ (Fig. 4B). The tuning curve demonstrates that the change in the dynamic range of a network due to a shift in E/I is a function of both the original, unperturbed state and the resulting change in $\kappa$.

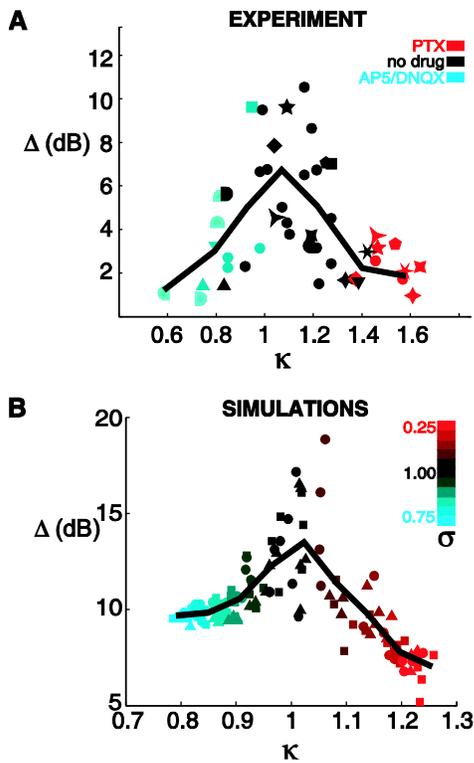

**FIGURE 4. The network tuning curve for dynamic range Δ near criticality.** *A*, In the experiments, Δ peaks close to κ ≅ 1 and drops rapidly with distance from the critical state. Paired measurements, where the normal (no-drug) condition was measured just before the drug condition, share the same symbol shape. *Circles*: unpaired measurement. *B*, In the model, Δ is also maximum for κ ≅ 1. Symbol shape indicates network size (*circles*: N = 250; *squares*: N = 500; *triangles*: N = 1000). Lines in *A* and *B* represent binned averages.

**Discussion**
We experimentally derived a tuning curve that linked the state of a cortical network with its ability to process inputs. When the network was closest to criticality, as indicated by neuronal avalanche dynamics, κ was close to one and the dynamic range was maximized. This is among the first experimental work to confirm theoretical predictions on the computational advantage of the critical state. Dynamic range has been predicted in simulations to peak in the critical state (Kinouchi and Copelli, 2006). Our simulations advance previous studies by 1) linking the dynamic range of a network with the spontaneous activity it generates and 2) making precise quantitative comparisons with experimental findings. Because the dynamic range increases with the ability of a network to map input differences into distinguishable network outputs, our result is also closely related to network-mediated separation, which has been predicted to peak in the critical state, at the transition from ordered to chaotic dynamics (Bertschinger and Natschlager, 2004;Legenstein and Maass, 2007). In contrast, an interesting result from our experiments and model is that variability of network response to a given stimulus intensity is highest in the critical state. Further investigation on the balance of reliability and variability in cortical networks is warranted.

Considering the simplicity of our model with all-to-all connectivity, absence of refractory period, and approximating inhibition by reducing σ, the overall agreement in the Δ−κ relationship between experiment and simulation is remarkable. The increase of variability in Δ as well as the drop in Δ due to deviation from κ = 1 was well matched between experiment and simulations. Such similarity supports the notion that universal principles are found in the critical state that are invariant to system specifics (Bak and Paczuski, 1995;Jensen, 1998;Stanley, 1971). The main quantitative difference was the lower Δ values for experiments compared to the model. Experimental noise, which is absent in the model, effectively adds a constant value to $S_{min}$ and $S_{max}$, which systematically reduces Δ.

Further neurophysiological insight into our results can be gained from Fig. 3. There it is shown that networks poorly discriminate small inputs in the hypo-excitable state, whereas they tend to saturate, failing to discriminate larger inputs in the hyper-excitable state. Both these reductions in performance result in reduced dynamic range compared to balanced networks. In line with these findings, dissociated cultures respond to inputs with a 'network spike' if σ > 1 (Eytan and Marom, 2006) and display a 'giant component' when transitioning to a hyperexcitable regime, which reduces the ability to discriminate inputs (Breskin et al., 2006). The balance of excitation and inhibition has been shown to be crucial for developing cortical circuits to accurately process sensory inputs (Hensch, 2005). Our results suggest that, functionally, the balance of excitation and inhibition is achieved when the dynamic range is maximized and cortical networks benefit from operating in the critical state.

*To the Journal of Neuroscience*
*Brief Communications*
*Section Cellular/Molecular*

**SUPPLEMENTAL MATERIAL**

# Neuronal Avalanches Imply Maximum Dynamic Range in Cortical Networks at Criticality

Woodrow L. Shew[1], Hongdian Yang[1,2], Thomas Petermann[1], Rajarshi Roy[2], and Dietmar Plenz[1]

### 1. Definition of spatiotemporal nLFP clusters

An example of the raw LFP recording from which we extract a nLFP cluster is shown in Figure S1. Each negative peak of each LFP fluctuation will be designated an nLFP with a time stamp and amplitude. A cluster of nLFPs was defined as a group of successive nLFPs on the array each separated at most by a time $\tau$. The threshold $\tau$ was chosen to separate the relatively short time scale of within-cluster activity from the longer time scale of quiet periods between clusters. Figure S2 shows a typical histogram of time intervals between nLFPs on the array (single experiment). *Red line*: choice of $\tau$.

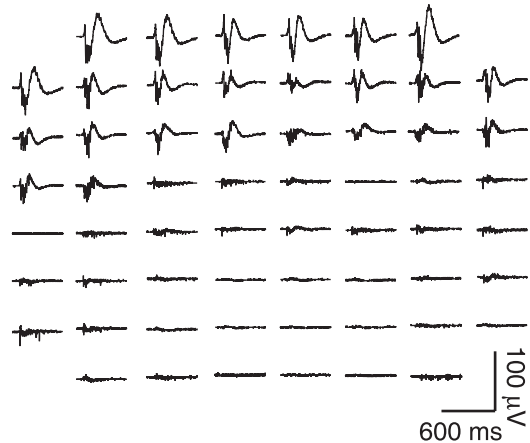

**FIGURE S1. Example population event.** Shown is an example of a population event revealed by widespread fluctuations in the local field potential (LFP) recorded by the micro-electrode array. Each trace is 60 ms of recorded LFP from one electrode in the array.

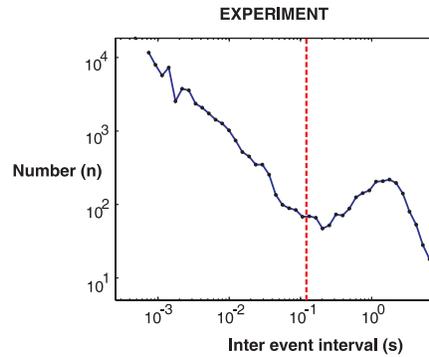

**FIGURE S2. Time threshold τ in burst definition.** Shown is an example of an inter-peak interval distribution of nLFPs on the array from one experiment. Two time scales are prominent: (1) short time intervals between peaks within periods of activity and (2) long time intervals reflecting periods with no activity (identified by the hump in this example). *Red line*: choice of τ.

## 2. Peri-stimulus time histograms of evoked activity

Although, not the focus of our study, evoked responses often exhibited complex temporal evolution. Shown in Figure S3 are typical peri-stimulus time histograms (PSTHs) of evoked activity for stimulus levels and three different drug conditions. The vertical axis represents average nLFP, normalized by the maximum observed nLFP. In the results presented in the main text, a response to a given stimulus level was quantified as the integral of the PSTH associated with that stimulus level. The three examples shown here were computed from the same data as the R(S) curves shown in Fig 3. of the main text. In the AP5/DNQX example, the PSTHs were mainly flat until a stimulus level of about 60 µA, demonstrating the insensitivity of the network when excitation is suppressed. At the other extreme, in the PTX example, the largest stimulus levels have similar PSTHs, demonstrating the saturation of the R(S) for large S when inhibition is suppressed.

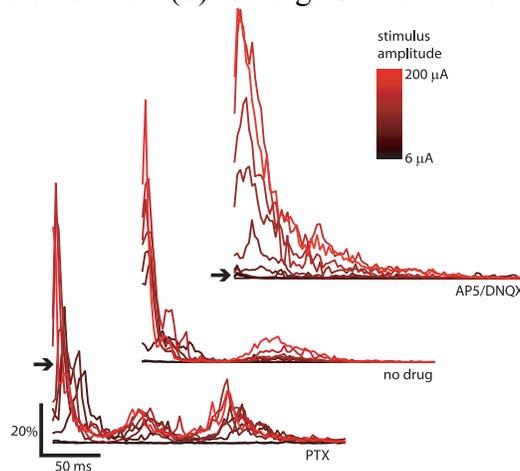

**FIGURE S3. Peri-stimulus time histograms (PSTHs) of evoked activity.** nLFP versus time averaged over 40 stimuli at each stimulus level (color coded) are compared for three drug conditions (left – PTX, middle – no drug, right – AP5/DNQX). In the AP5/DNQX condition the system is relatively insensitive due to suppressed excitation, i.e. the PSTH is flat until a stimulus level of about 60 µA is reached. In the PTX condition, the largest stimulus levels result in very similar PSTHs demonstrating the tendency for response to saturate when inhibition is suppressed.   Note that the response to a given stimulus level, R, in the main text was defined as the integral of the PSTH.

## 3.  Quantification of the network state using kappa

In previous work by our lab (Beggs and Plenz, 2003), we used a different measure than κ for characterizing the spontaneous activity of the system. Here we discuss our motivations for this change. In previous work, an estimate of the parameter σ was computed from experimental data with a method based on counting the number of electrodes which recorded negative LFP peaks within successive time periods. The ratio of the active electrode count in the second time period to the count in the first time period was taken as the σ estimate.  This method yielded σ ≅ 1 during neuronal avalanches as expected from theory. Later, this measure was found to give unexpected results for apparently supercritical states (Plenz, 2005). To better understand these observations, we tested the method for estimating σ using our model, where the true σ is known and exact counting of numbers of active sites is feasible (Figure S4A).  Using time periods with various durations and starting times within a cluster, we found that the previously used method for estimating σ was very accurate when the network was critical.  However, away from the critical state, this method was highly sensitive to the temporal resolution used. This may explain why the previous method robustly and correctly estimated σ ≅ 1 for neuronal avalanches, but was unreliable for supercritical states.

   As a measure of the system state, κ avoids the above discussed difficulties primarily because it does not depend on precise temporal resolution.  The comparison of κ with σ (Figure S4A) demonstrates its superiority over previous methods of estimating σ.  We tried several alternative definitions for κ, which all outperformed the previous method of estimating σ and are compared to each other in figure 2SB.  Our choice of definition was guided by the aim to make the match between κ and σ as close as possible using the model data.  Our use of cumulative distributions rather than the PDFs to calculate κ avoids sensitivity to binning choices, which are necessary for constructing a PDF. κ is in the same family of nonparametric comparisons of cumulative distributions as the Kolmogorov-Smirnov (KS) test and the Kuiper's test.  The KS statistic is the single maximum difference between two cumulative distributions and Kuiper's test is the sum of the absolute values of the maximum positive difference and the maximum negative difference.  Compared to Kuiper's test, κ simply takes the sum of more than two differences without the absolute value.  We note that the Kolmogorov-Smirnov statistic does a rather poor job for our purposes (Fig. S4B, *green*). Furthermore, our choice of logarithmic spacing of the $\beta_k$ values provides a more linear relationship between κ and σ compared to a linear spacing of $\beta_k$ (Fig. S4B, *blue*).

   Finally, we point out that several aspects of the definition are very robust. For example, if we alter the upper end of the range of burst sizes used to generate the reference CDF, from $10^3$ to $10^5$ μV, the measured κ values change only slightly (data not shown).  We note that the typical maximum burst size measured from a given culture also ranged from $10^3$ to $10^5$ μV.  Furthermore, κ is nearly unchanged with respect to the number of differences computed between the two CDFs for all *m* >5 (*m* as defined in the main text; *data not shown*). However, since it is a statistical measure, κ is naturally more prone to error when sample sizes are low. To account for this, we only included experiments in our analysis in which we observed at least 200 spontaneous nLFP clusters.

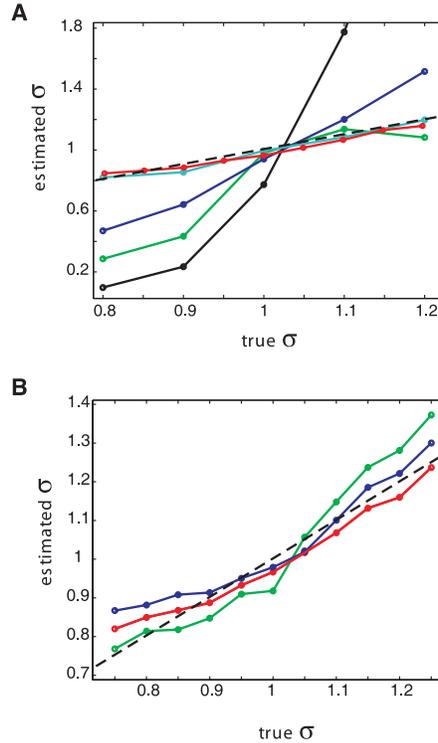

**FIGURE S4. Estimating σ using κ.** *A*, Using our model, we compared the true value of σ to an estimate based on activity clusters. The black dashed line has slope 1, representing a perfect estimate. The new measure κ (*red*) accurately estimated the underlying branching parameter over a wide range of subcritical to supercritical states. In contrast, an estimate of the branching parameter σ based on calculating ratios of descendants to ancestors, i.e. active sites during period *t+1* divided by active sites during period *t*, varied significantly in its precision depending on network state and time periods used. A reliable estimate was achieved when the correct temporal resolution for each network state was available and when comparing the first two consecutive time steps of a cluster (*light blue*; $\sigma_{est} = A(2)/A(1)$; $A(t1)$ and $A(t2)$ are the number of active neurons during time *t1* and t2 respectively). Importantly, when the exact temporal resolution was not known, estimates tended to stray widely from the real value for subcritical and supercritical dynamics. For $\sigma_{est} = A(11:20)/A(1:10)$ (*black*) and subcritical dynamics, clusters tended to die during t1 = 1:10, leading to an underestimate of σ. Conversely, clusters tended to expand supralinearly for t2=11:20 in supercritical dynamics, leading to an overestimate of σ. For comparison, other variations of sampling situations were also plotted (*green*: $\sigma_{est} = A(29:31)/A(26:28)$; *dark blue*: $\sigma_{est} = A(9:11)/A(6:8)$).

*B*, Comparison in the accuracy of estimating true σ using various metrics to quantify differences between CDF (network model). *Red*: κ with 10 logarithmically spaced $\beta_k$ values provides the most reliable and linear estimate of σ. *Blue*: modified κ with linear spacing of $\beta_k$ values reveals increased mismatch for extreme sub-critical and supercritical dynamics. *Green*: Kolmogorov-Smirnov statistic using the maximal distance between two CDF performed worst.

## 4. Robustness of results to variations in maximal stimulus amplitude

The maximum stimulus amplitude, 200 µA, in the experiments was chosen to maximize the stimulus range without damaging the tissue. With this range of stimuli, the response curves did not always saturate. To test whether this limitation impacts our hypothesis, that is $\Delta$ is maximized for $\kappa = 1$, we recalculated the $\Delta$ vs. $\kappa$ curve with deliberately truncated stimulus ranges in both the model and the experiment. In figure S5A we demonstrate that our data support our hypothesis even for the limited stimulus size range available to us experimentally. Only for an extremely shortened stimulus range does the hypothesis become non-testable as shown in the corresponding model simulations (Fig. S5B).

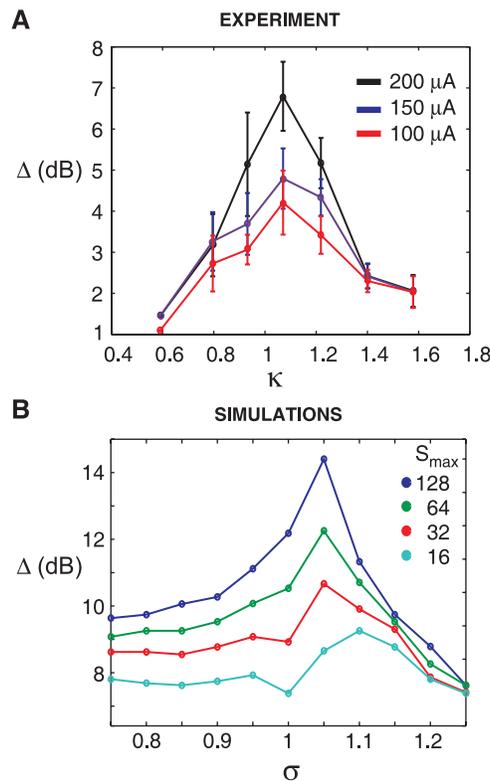

**FIGURE S5. Effect of limited stimulus range on Δ.** *A*, The *blue line* is a re-plot of binned and averaged data from figure 4A of the main text. The green and red lines represent the same experiments, but reprocessed using only <150 and <100 µA respectively. The peak of $\Delta$ near $\kappa=1$ is attenuated, but still exists. *B*, In the model, we verify that we should expect attenuation of the $\Delta(\kappa)$ curve, when the stimulation range is decreased. The strong peak vanishes only for a severely truncated range ($S_{max}=16$).

## 5. Robustness of simulation results to network size

Finally, we tested the effects of changing the number of neurons in the model. We found that for a fixed range of stimulus intensities, i.e. number of initially active sites, the $\Delta(\sigma)$ curve was largely unchanged. For $\sigma<1$, there was a tendency for larger systems to have decreased $\Delta$. Each point on the curves in figure S6 is an average over 6 different simulations (same data as shown in Fig. 4B of the main text)

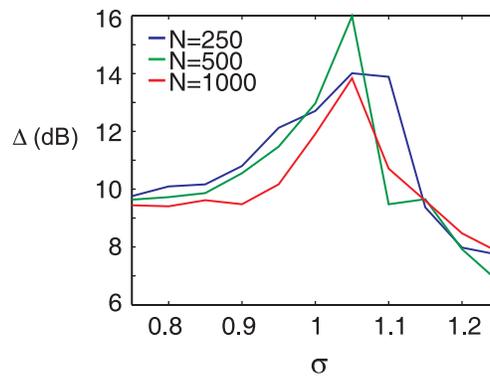

**FIGURE S6. Effect of network size on Δ.** Increasing the system size from *N*=250 to *N*=1000 model neurons causes only slight shifts in Δ. For σ < 1 there is a tendency for slightly lower Δ at higher *N*.